\documentclass[
english,
10pt,%
notitlepage,%
twocolumn,
nofootinbib,% В текущей версии REVTeX c этой опцией
superscriptaddress,%
nolongbibliography,
noshowpacs,%
centertags]%
{revtex4-1}

\usepackage{rotating, amssymb}
\usepackage{graphicx, caption2}

\newcommand{\Teff}{$T_{eff}$}
\newcommand{\lgg}{\rm{lg}~$g$}
\newcommand{\Vt}{$V_t$} 

\newcommand{\kms}{$km\,s^{-1}$}
\def\refitem#1{\relax}

\sloppypar

\begin{document}

\title{\bf Chemical Composition of the Atmospheres of Red Giants
with High Space Velocities}

\author{\bf Yu. V. Pakhomov}
\affiliation{{\it Institute of Astronomy, Russian Academy of Sciences,
Pyatnitskaya ul. 48, Moscow, 109017 Russia}, pakhomov@inasan.ru}

\begin{abstract}
\noindent
The results of a comparative analysis of the elemental abundances in the atmospheres of 14 red 
giants with high Galactic space velocities are presented. For almost all of the chemical elements considered, 
the their abundance trends with metallicity correspond to those constructed for thick-disk dwarfs. In the 
case of sodium, the main factor affecting the [Na/Fe] abundance in the stellar atmosphere for red giants is 
the surface gravity that characterizes the degree of development of the convective envelope. The difference 
between the [Na/Fe] abundances in the atmospheres of thin-and thick-disk red giants has been confirmed. 

{\bf Keywords:} stellar spectroscopy, stellar atmospheres, red giants,
stellar evolution, kinematics, Galactic chemical evolution.

\noindent
{\bf PACS: } 97.10.Cv, 97.10.Ex, 97.10.Ri, 97.10.Tk, 97.20.Li,
98.35.Bd, 98.35.Df, 98.35.Pr
\end{abstract}

\maketitle

\section*{INTRODUCTION}
\noindent

Our Galaxy is currently believed to be a set of 
subsystems that differ by their sizes, locations in the 
Galaxy, characteristic stellar velocities, and stellar 
populations. The best-known subsystems are the 
nucleus, the bulge, the thick and thin disks, and the 
halo. Each subsystem can contain dynamical stellar 
groups or streams (see, e.g., \cite{1958MNRAS.118...65E},
\cite{1958MNRAS.118..154E}). 
At a Galactocentric distance of about 8.5 kpc, where 
the Solar system is located, there are stars of at 
least three Galactic subsystems: the thick and thin 
disks and the halo. In addition, stars with significant 
orbital eccentricities that can belong to the bulge 
exist in the inner Galaxy. Such stars are characterized
 by an enhanced metallicity, and super-metalrich
 stars are more often observed among the bulge 
objects \cite{2009ARep...53..685P}. In contrast, the 
halo stars exhibit a low metallicity and a large scatter 
of orbital inclinations and eccentricities. Thus, the 
structure and chemical evolution of the Galaxy can 
be investigated by studying the motion of stars in 
the solar neighborhood, their physical characteristics, 
and the abundances of chemical elements in their 
atmospheres. 

The chemical composition of the stellar atmospheres
 changes as the star on the temperature. 
luminosity diagram passes from the main sequence 
to the red giant branch in the course of its evolution. 
The hydrogen burning reactions cease in the core and 
begin in the shell around the core. At this stage of 
its life, the structure of the star changes significantly: 
the star rapidly expands, and an extensive convective 
shell that provides the first deep mixing of stellar 
matter is formed. The nuclear burning products are 
brought into the stellar atmosphere and become observable.
 The changes in the CNO abundances are 
well known: carbon and oxygen are reduced approximately
 by 0.1-0.3 dex, while nitrogen is enhanced 
by 0.1-0.4 dex. In addition to these elements, the 
changes can also affect other elements. 

In this paper, we analyze the chemical composition
 of red giants with Galactic velocities that exceed 
those typical of thin-disk stars in the solar neighborhood
 and the main causes of the changes in elemental
 abundances. These changes reflect not only 
the chemical evolution of the Galactic subsystem to 
which they belong but also the stellar evolution processes.

\section*{OBSERVATIONS}
\noindent

\begin{table}
\captionstyle{flushleft}
\onelinecaptionsfalse 
\centering
\caption{The list of studied stars with the membership probabilities ($p$) in some 
Galactic subsystems}
\label{tab:list}
\begin{tabular}{|c|c|c|c|c|c|c|c|c|}
\hline
N&HD    &$\alpha_{2000}$&$\delta_{2000}$&$m_V$&SpType&\multicolumn{3}{c|}{$p$,
\%} \\
&& &&  & &thin&Hercules& thick\\
&& h:m:s&$\circ:\prime:\prime\prime$&  &  &disc&stream&disc\\ 
\hline
 1&  2901&00:32:47.6&+54:07:11& 6.91& K2III    &   - &   - &  99\\
 2& 10550&01:42:43.5&-03:41:24& 4.98& K3II-III &  27 &  58 &  13\\
 3& 92095&10:39:05.7&+53:40:05& 5.55& K3III    &  16 &   - &  83\\
 4& 94600&10:55:44.4&+33:30:25& 5.02& K1III    &  91 &   - &   8\\
 5& 94669&10:56:14.5&+42:00:29& 6.03& K2III    &  30 &  41 &  27\\
 6& 94860&10:59:56.8&+77:46:12& 6.18& G9III    &  15 &   4 &  80\\
 7& 99902&11:29:41.9&+15:24:47& 5.74& K4III    &  87 &   - &  11\\
 8&100696&11:36:02.7&+69:19:22& 5.19& K0III    &  52 &   - &  47\\
 9&104985&12:05:15.1&+76:54:20& 5.78& G9III    &  38 &   - &  61\\
10&127243&14:28:37.8&+49:50:41& 5.58& G3IV     &   - &   - &  99\\
11&141353&15:48:13.3&+13:47:21& 5.98& K2III    &   1 &   1 &  97\\
12&141472&15:46:34.7&+55:28:29& 5.94& K3III    &   - &   - &  99\\
13&152879&16:55:22.2&+18:25:59& 5.35& K4III    &   4 &   - &  95\\
14&197752&20:44:52.5&+25:16:14& 4.92& K2III    &  41 &   - &  58\\
\hline
\end{tabular}
\end{table}

We selected the stars for our observations based on 
their Galactic velocities calculated from reduced Hipparcos
 parallaxes \cite{2007A&A...474..653V} and 
CORAVEL radial velocities \cite{1987A&AS...67..423M}. 
We chose the red giants for which at least one of 
the Galactic velocity vector components (UVW) 
exceeded its typical values for the thin disk (34.5, 
22.5, 18.0)~\kms~\cite{2007A&A...474..653V}. The 
effective temperatures \Teff\ and surface gravities \lgg\ 
for the program stars were preestimated from photometric data and reduced Hipparcos
parallaxes~\cite{2007A&A...474..653V}. Subsequently, the stars with 
effective temperatures \Teff\ from 4000 to 5200~K and 
surface gravities \lgg\ from 0.5 to 3.0 were chosen 
from the list. The list of program stars is presented in 
Table~\ref{tab:list}. 

The spectroscopic observations of the chosen 
stars were performed in 2007-2008 with a two-arc 
echelle spectrograph attached to a 2.16-m telescope 
at the Xinglong station of the National Astronomical 
Observatories of China (NAOC). The spectrograph 
operated in the red-band mode. The detector was a 
1024x1024 CCD array on which 40 spectral orders 
in the range from 5565 to 9194~\AA\ were recorded. The 
spectrograph resolution was $R$=40\,000; the signal-to-noise 
ratio in the spectra was $S/N>150$. 

The echelle package of the MIDAS software system
 was used for the spectroscopic data reduction, 
the search for and extraction of the spectral orders, 
the wavelength calibration using the spectrum of a 
thorium-argon lamp, and the spectrum normalization.
 The equivalent widths of the selected spectral 
lines were measured in the EW code that I wrote. 

\section*{DETERMINING THE PARAMETERS OF THE STELLAR ATMOSPHERES}

\begin{table}
\captionstyle{flushleft}
\onelinecaptionsfalse 
\centering
\caption{Atmospheric parameters of the investigated stars, their physical parameters, and interstellar extinctions}
\label{tab:param}
\begin{tabular}{|c|c|c|c|c|c|c|c|c|c|c|c|c|}
\hline
N&  HD   &\Teff &\lgg &\Vt &[Fe/H]&  Mass  & Age&$A_V$ \\
 &       & K    &     &\kms&      &$M_\odot$&lg, years& m \\
\hline
1 &  2901 & 4350 &2.15 &1.25&-0.46 & 0.7$\pm$0.3& 9.4$\pm$0.5& 0.1    \\
2 & 10550 & 4290 &1.25 &1.50& 0.05 &   7$\pm$1  & 7.7$\pm$0.4& 0.1    \\
3 & 92095 & 4430 &2.15 &1.38&-0.02 & 1.9$\pm$0.3& 9.0$\pm$0.3& $<$0.1   \\
4 & 94600 & 4660 &2.50 &1.20&-0.05 & 1.7$\pm$0.3& 9.0$\pm$0.3&   0    \\
5 & 94669 & 4620 &2.65 &1.25& 0.00 & 1.5$\pm$0.4& 8.9$\pm$0.2&   0    \\
6 & 94860 & 4970 &2.75 &1.20&-0.03 & 2.8$\pm$0.3& 8.7$\pm$0.2& $<$0.1   \\
7 & 99902 & 4380 &2.25 &1.30& 0.22 & 1.9$\pm$0.5& 9.0$\pm$0.3& $<$0.1   \\
8 &100696 & 4920 &2.70 &1.26&-0.21 & 2.4$\pm$0.3& 8.6$\pm$0.3&  0     \\
9 &104985 & 4830 &2.80 &1.28&-0.10 & 1.9$\pm$0.3& 9.0$\pm$0.3&  0     \\
10&127243 & 5100 &1.75 &1.28&-0.44 & 1.6$\pm$0.3& 9.1$\pm$0.3&  0     \\
11&141353 & 4280 &1.95 &1.37&-0.09 & 2.0$\pm$0.4& 9.0$\pm$0.4& $<$0.1   \\
12&141472 & 4180 &1.50 &1.40&-0.27 & 1.1$\pm$0.3& 9.0$\pm$0.5& $<$0.1   \\
13&152879 & 4170 &1.80 &1.39& 0.02 & 1.4$\pm$0.5& 9.5$\pm$0.5& $<$0.1   \\
14&197752 & 4570 &2.25 &1.33& 0.03 & 2.6$\pm$0.5& 8.8$\pm$0.5& $<$0.1   \\
\hline
\end{tabular}
\end{table}

%\clearpage
\begin{table*}
\caption{Elemental abundances in the atmospheres of the program stars}
\label{tab:abund}
\begin{tabular}{|c|c|c|c|c|c|c|c|c|c|c|c|c|c|c|}
\hline
      &\multicolumn{2}{c|}{HD 2901}&\multicolumn{2}{c|}{HD
10550}&\multicolumn{2}{c|}{HD 92095}&\multicolumn{2}{c|}{HD
94600}&\multicolumn{2}{c|}{HD 94669}&\multicolumn{2}{c|}{HD
94860}&\multicolumn{2}{c|}{HD 99902}\\
\hline
     &  N  &   [El/H]       &  N  &   [El/H]       &  N  &   [El/H]       &  N 
&   [El/H]       &  N  &   [El/H]       &  N  &   [El/H]       &  N  &   [El/H] 
 \\
\hline
 NaI &   2 &  -0.39$\pm$0.04&   2 &   0.67$\pm$0.08&   2 &  -0.04$\pm$0.04&   2
&  -0.04$\pm$0.07&   2 &   0.00$\pm$0.04&   2 &   0.27$\pm$0.02&   2 &  
0.40$\pm$0.04\\
 MgI &   2 &  -0.12$\pm$0.01&   2 &   0.16$\pm$0.02&   2 &   0.06$\pm$0.02&   2
&   0.00$\pm$0.00&   2 &   0.04$\pm$0.01&   2 &   0.24$\pm$0.02&   2 &  
0.30$\pm$0.08\\
 AlI &   2 &  -0.12$\pm$0.04&   2 &   0.21$\pm$0.07&   2 &   0.04$\pm$0.03&   2
&  -0.05$\pm$0.00&   2 &   0.16$\pm$0.08&   2 &   0.26$\pm$0.01&   2 &  
0.32$\pm$0.05\\
 SiI &   8 &  -0.18$\pm$0.07&   2 &   0.04$\pm$0.01&   7 &   0.06$\pm$0.06&  10
&  -0.02$\pm$0.07&   9 &   0.06$\pm$0.08&   8 &   0.10$\pm$0.03&   8 &  
0.29$\pm$0.08\\
 CaI &   3 &  -0.22$\pm$0.07&  -- &                &   3 &   0.02$\pm$0.07&   4
&   0.11$\pm$0.12&   3 &   0.13$\pm$0.06&   4 &   0.13$\pm$0.02&   3 &  
0.29$\pm$0.06\\
 ScI & --  &    --          & --  &    --          &   1 &   0.03         &   2
&  -0.04$\pm$0.14&   2 &   0.24$\pm$0.14&   1 &   0.30         & --  &    --    
     \\
 ScII&   5 &  -0.12$\pm$0.06&   2 &   0.06$\pm$0.03&   4 &   0.18$\pm$0.05&   3
&   0.00$\pm$0.03&   5 &   0.15$\pm$0.06&   3 &   0.39$\pm$0.03&   4 &  
0.28$\pm$0.10\\
 TiI &  20 &  -0.18$\pm$0.07&   5 &   0.01$\pm$0.03&  20 &  -0.05$\pm$0.06&  25
&  -0.07$\pm$0.06&  24 &   0.01$\pm$0.08&  24 &   0.18$\pm$0.05&  14 &  
0.24$\pm$0.06\\
 VI  &   7 &  -0.13$\pm$0.07&   2 &  -0.04$\pm$0.01&   7 &   0.10$\pm$0.08&  20
&   0.00$\pm$0.06&  13 &   0.15$\pm$0.07&  25 &   0.18$\pm$0.07&   3 &  
0.21$\pm$0.07\\
 CrI &   6 &  -0.46$\pm$0.07&   4 &  -0.03$\pm$0.08&   7 &  -0.11$\pm$0.09&  11
&  -0.10$\pm$0.09&  10 &   0.00$\pm$0.07&   9 &  -0.05$\pm$0.05&   8 &  
0.15$\pm$0.09\\
 MnI &   1 &  -0.79         &   1 &  -0.17         &   1 &  -0.29         &   1
&  -0.31         &   1 &  -0.19         & --  &    --          &   1 &   0.03   
     \\
 FeI &  66 &  -0.46$\pm$0.06&  16 &   0.05$\pm$0.07&  58 &  -0.02$\pm$0.06&  73
&  -0.05$\pm$0.06&  63 &   0.00$\pm$0.07&  66 &  -0.03$\pm$0.06&  55 &  
0.22$\pm$0.07\\
 FeII&   6 &  -0.48$\pm$0.10&   4 &   0.00$\pm$0.07&   6 &  -0.07$\pm$0.10&   5
&  -0.07$\pm$0.08&   5 &   0.00$\pm$0.06&   5 &  -0.07$\pm$0.07&   2 &  
0.13$\pm$0.03\\
 CoI &   4 &  -0.40$\pm$0.08&   4 &   0.05$\pm$0.09&   4 &   0.04$\pm$0.08&   5
&  -0.11$\pm$0.07&   5 &   0.04$\pm$0.07&   5 &   0.07$\pm$0.12&   8 &  
0.27$\pm$0.09\\
 NiI &  21 &  -0.40$\pm$0.06&   6 &   0.01$\pm$0.05&  13 &  -0.03$\pm$0.06&  18
&  -0.09$\pm$0.05&  14 &   0.03$\pm$0.07&  18 &   0.03$\pm$0.07&  14 &  
0.23$\pm$0.06\\
 YII & --  &    --          & --  &    --          & --  &    --          & -- 
&    --          & --  &    --          &   1 &  -0.32         &   2 &  
0.14$\pm$0.03\\
 ZrII& --  &    --          &   1 &  -0.06         &   1 &   0.09         &   1
&   0.23         &   1 &   0.23         & --  &    --          &   1 &   0.18   
     \\
 MoI &   2 &  -0.42$\pm$0.02& --  &    --          &   1 &  -0.31         & -- 
&    --          & --  &    --          &   1 &  -0.02         &   1 &   0.06   
     \\
 BaII&  1  &  -0.15         & --  &    --          &   1 &   0.56         &   1
&   0.15         &  1  &    0.53        &   1 &   0.06         & --  &    --  
 
     \\
 LaII& --  &    --          &   1 &   0.07         & --  &    --          & -- 
&    --          & --  &    --          & --  &    --          &   1 &   0.15   
     \\
 CeII&   1 &  -0.45         & --  &    --          &   1 &   0.79         &   1
&   0.37         &   1 &   0.83         &   1 &   0.24         &   1 &   0.39   
     \\
 NdII&   4 &  -0.29$\pm$0.06&   1 &   0.07         &   3 &   0.22$\pm$0.06&   2
&   0.17$\pm$0.10&   1 &  -0.03         & --  &    --          &   2 &  
0.42$\pm$0.02\\
 EuII&   1 &   0.03         &   1 &   0.13         &   1 &   0.10         &   1
&   0.06         &   1 &   0.15         &   1 &   0.28         &   1 &   0.37   
     \\
\hline
\hline
      &\multicolumn{2}{c|}{HD 100696}&\multicolumn{2}{c|}{HD
104985}&\multicolumn{2}{c|}{HD 127243}&\multicolumn{2}{c|}{HD
141353}&\multicolumn{2}{c|}{HD 141472}&\multicolumn{2}{c|}{HD
152879}&\multicolumn{2}{c|}{HD 197752}\\
\hline
     &  N  &   [El/H]       &  N  &   [El/H]       &  N  &   [El/H]       &  N 
&   [El/H]       &  N  &   [El/H]       &  N  &   [El/H]       &  N  &   [El/H] 
 \\
\hline
 NaI &   2 &  -0.21$\pm$0.03&   2 &  -0.10$\pm$0.02&   3 &  -0.58$\pm$0.03&   2
&  -0.03$\pm$0.03&   2 &  -0.31$\pm$0.02&   2 &   0.28$\pm$0.07&   2 &  
0.17$\pm$0.04\\
 MgI &   2 &  -0.10$\pm$0.02&   2 &   0.11$\pm$0.00&   2 &  -0.15$\pm$0.03&   2
&   0.18$\pm$0.01&   2 &  -0.03$\pm$0.01&   2 &   0.20$\pm$0.09&   2 &  
0.15$\pm$0.03\\
 AlI &   2 &  -0.12$\pm$0.01&   2 &   0.12$\pm$0.05&   2 &  -0.26$\pm$0.02&   2
&   0.10$\pm$0.09&   2 &  -0.03$\pm$0.12&   1 &  -0.02         &   2 &  
0.12$\pm$0.08\\
 SiI &  10 &  -0.10$\pm$0.08&  13 &   0.08$\pm$0.05&  10 &  -0.29$\pm$0.07&   8
&   0.13$\pm$0.07&   8 &  -0.13$\pm$0.06&   7 &   0.13$\pm$0.09&   6 &  
0.05$\pm$0.04\\
 CaI &   4 &  -0.04$\pm$0.04&   3 &   0.06$\pm$0.04&   4 &  -0.25$\pm$0.05&   3
&  -0.04$\pm$0.07&   4 &  -0.17$\pm$0.05&   2 &   0.06$\pm$0.00&   3 &  
0.13$\pm$0.05\\
 ScI &   2 &  -0.17$\pm$0.11&   1 &   0.15         &   1 &  -0.67         & -- 
&    --          &   1 &  -0.14         & --  &    --          &   1 &   0.05   
     \\
 ScII&   4 &  -0.14$\pm$0.07&   5 &   0.25$\pm$0.07&   3 &  -0.37$\pm$0.07&   5
&  -0.02$\pm$0.09&   5 &  -0.08$\pm$0.07&   3 &   0.02$\pm$0.11&   6 &  
0.08$\pm$0.04\\
 TiI &  24 &  -0.19$\pm$0.06&  20 &   0.03$\pm$0.07&  15 &  -0.27$\pm$0.07&  17
&  -0.07$\pm$0.08&  13 &  -0.15$\pm$0.05&   9 &   0.02$\pm$0.08&  20 &  
0.00$\pm$0.06\\
 VI  &  26 &  -0.20$\pm$0.08&  18 &   0.07$\pm$0.07&   8 &  -0.44$\pm$0.02&   3
&  -0.16$\pm$0.05&   3 &  -0.11$\pm$0.04&   1 &   0.30         &   6 &  
0.15$\pm$0.05\\
 CrI &   8 &  -0.21$\pm$0.08&   7 &  -0.13$\pm$0.05&   1 &  -0.56         &  11
&  -0.19$\pm$0.11&   4 &  -0.31$\pm$0.04&   6 &   0.00$\pm$0.07&  11 & 
-0.02$\pm$0.04\\
 MnI &   1 &  -0.38         &   1 &  -0.53         &   1 &  -0.84         &   1
&  -0.36         &   1 &  -0.81         & --  &    --          &   1 &  -0.20   
     \\
 FeI &  79 &  -0.21$\pm$0.06&  69 &  -0.10$\pm$0.06&  53 &  -0.51$\pm$0.08&  57
&  -0.09$\pm$0.07&  51 &  -0.33$\pm$0.08&  38 &   0.02$\pm$0.05&  62 &  
0.03$\pm$0.05\\
 FeII&   6 &  -0.20$\pm$0.08&   5 &  -0.12$\pm$0.08&   6 &  -0.57$\pm$0.07&   4
&  -0.11$\pm$0.07&   3 &  -0.44$\pm$0.09&   2 &   0.00$\pm$0.11&   4 &  
0.01$\pm$0.04\\
 CoI &   4 &  -0.27$\pm$0.05&   5 &  -0.02$\pm$0.07&   2 &  -0.39$\pm$0.00&   6
&   0.10$\pm$0.05&   6 &  -0.26$\pm$0.07&   4 &   0.12$\pm$0.07&   5 & 
-0.04$\pm$0.08\\
 NiI &  25 &  -0.22$\pm$0.07&  20 &  -0.06$\pm$0.05&  13 &  -0.50$\pm$0.07&  11
&  -0.07$\pm$0.05&   9 &  -0.30$\pm$0.05&   6 &  -0.03$\pm$0.07&  15 &  
0.03$\pm$0.06\\
 YII &   1 &  -0.15         & --  &    --          & --  &    --          &   1
&  -0.25         &   2 &  -0.21$\pm$0.02&   1 &   1.20         &   1 &  -0.02   
     \\
 ZrII&   1 &   0.03         & --  &    --          & --  &    --          &   1
&  -0.09         &   1 &  -0.06         & --  &    --          &   1 &  -0.06   
     \\
 BaII&   1 &   0.09         &   1 &   0.14         &   1 &  -0.36         &   1
&   0.14         & --  &    --          & --  &    --          &   1 &   0.29   
     \\
 CeII& --  &    --          & --  &    --          & --  &    --          & -- 
&    --          &   1 &   0.16         & --  &    --          &   1 &   0.47   
     \\
 NdII&   2 &   0.10$\pm$0.14&   1 &   0.04         & --  &    --          &   2
&   0.01$\pm$0.10&   3 &   0.03$\pm$0.09&   2 &   0.09$\pm$0.10&   2 &  
0.10$\pm$0.06\\
 EuII&   1 &  -0.08         &   1 &   0.20         &   1 &  -0.31         &   1
&   0.09         &   1 &  -0.03         &   1 &   0.09         &   1 &   0.12   
     \\
\hline
\end{tabular}
\end{table*}

\begin{figure}[h]
\captionstyle{flushleft}
\onelinecaptionsfalse 
\centering
\includegraphics[width=7cm]{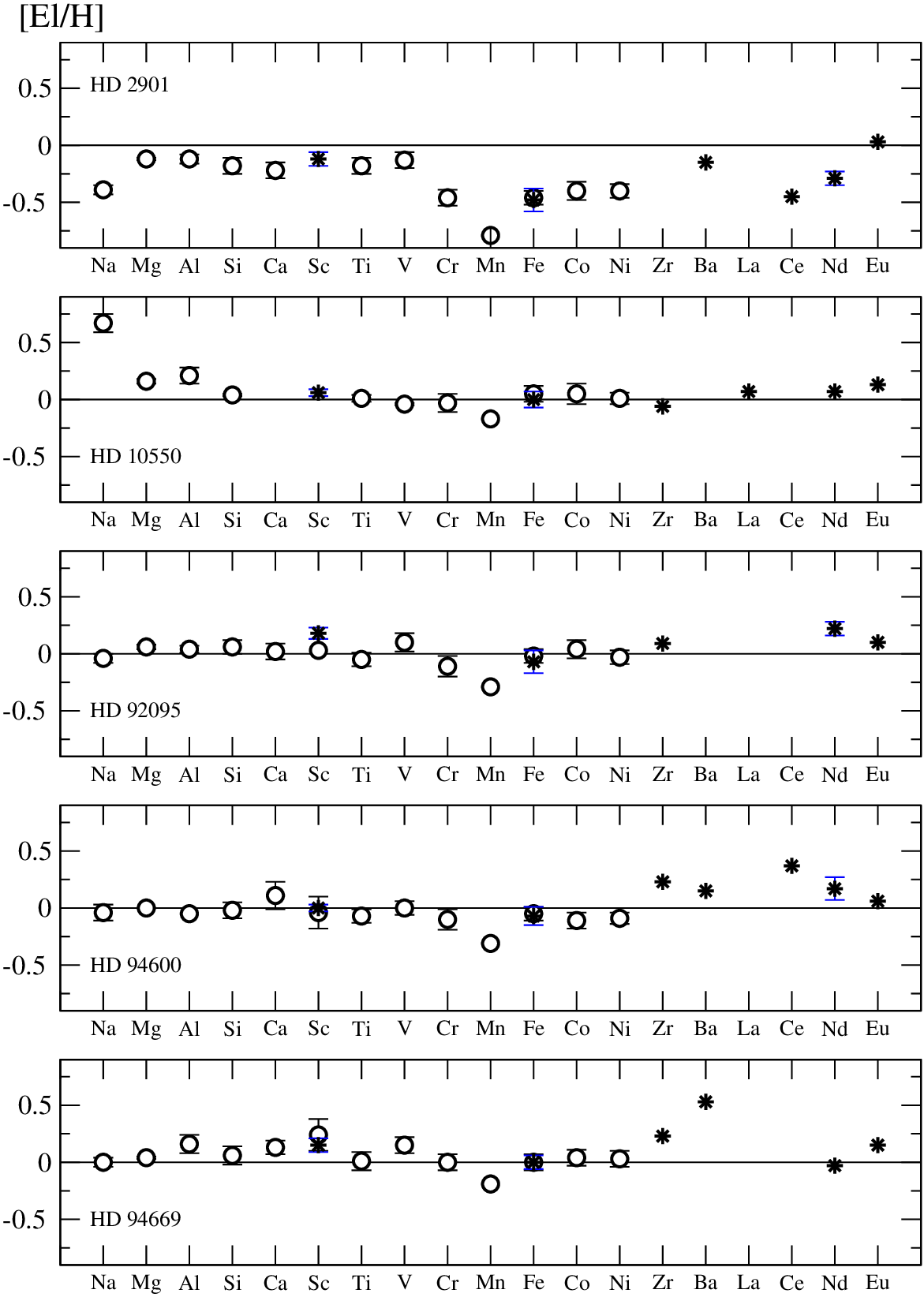}
\caption{\rm Relative elemental abundances in the atmospheres of the program stars. The open circles and asterisks denote the 
abundances determined from spectral lines of neutral and ionized atoms, respectively. }
\label{fig:abund}
\end{figure}

\begin{figure}[h]
\centering
\captionstyle{flushleft}
\onelinecaptionsfalse 
\includegraphics[width=7cm]{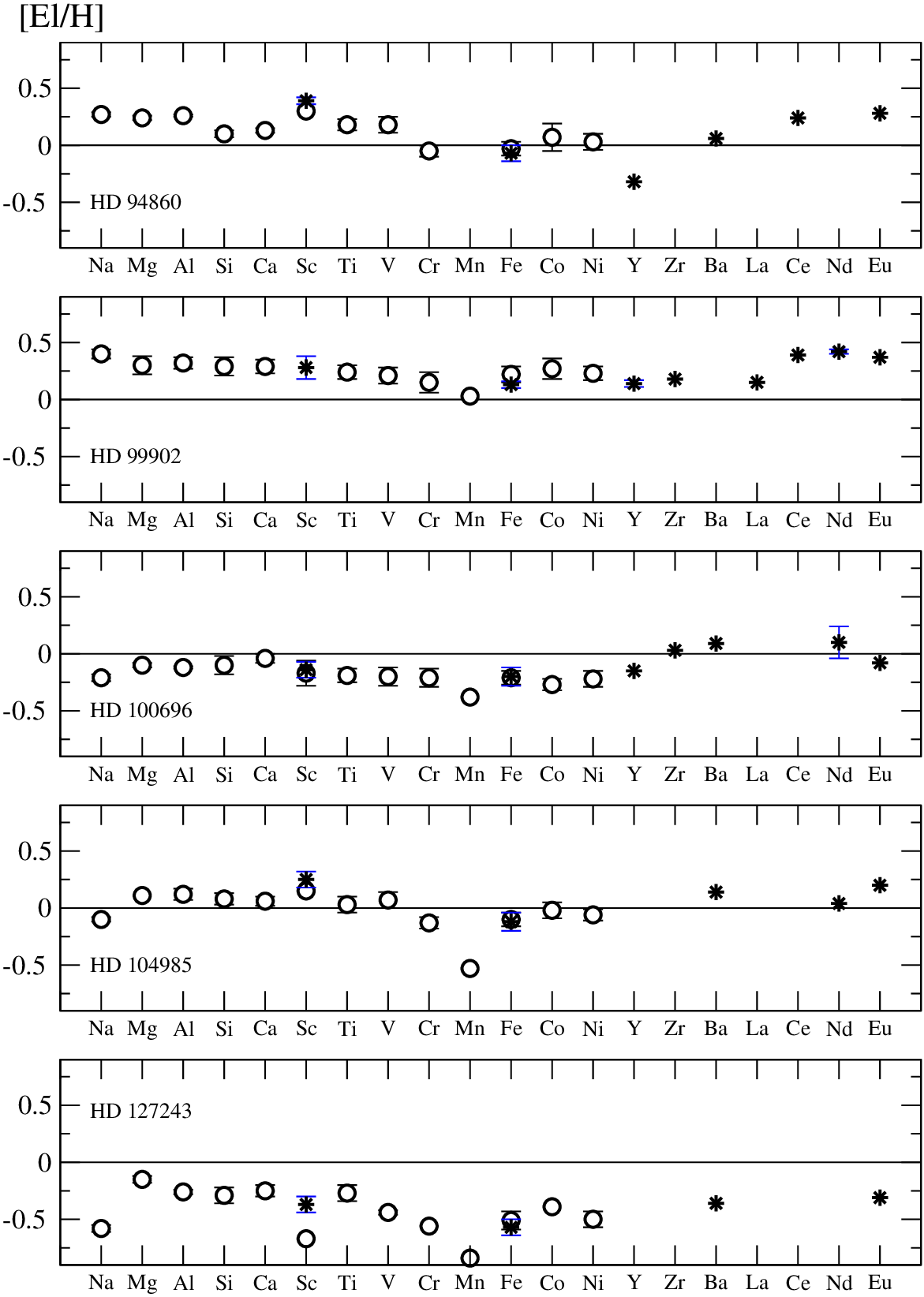}
\includegraphics[width=7cm]{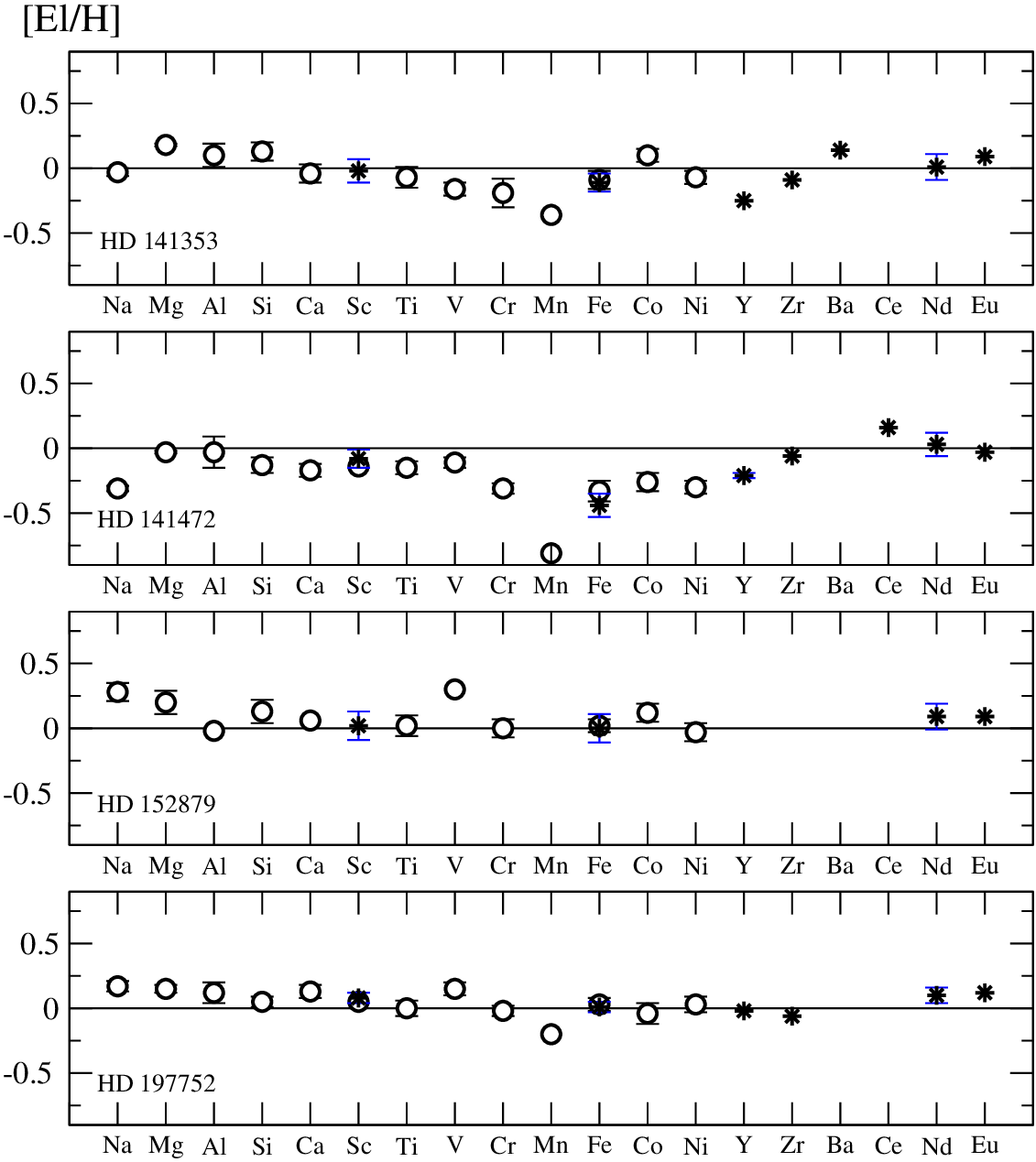}
\addtocounter{figure}{-1}
\caption{\rm Relative elemental abundances in the atmospheres of the program stars. The open circles and asterisks denote the 
abundances determined from spectral lines of neutral and ionized atoms, respectively. 
(continue)}
\end{figure}

\begin{table}
\captionstyle{flushleft}
\onelinecaptionsfalse 
\centering
\caption{Changes of the elemental abundances in the atmospheres of the stars
HD~100673 (\Teff=4920~K \lgg=2.70 \Vt=1.26~\kms) и HD~152879 (\Teff=4170K \lgg=1.80
\Vt=1.39~\kms) when changing the model parameters by $\Delta$\Teff=+100~K,
$\Delta$\lgg=+0.10, $\Delta$\Vt=+0.10~\kms\ and the total change in abundance
$\Delta$}
\label{tab:abnerr}
\begin{tabular}{|c|r|r|r|r|r||r|r|r|r|r|}
\hline
Elem&
N&\multicolumn{4}{c||}{$\Delta[El/H]_{HD100673}$}&
N&\multicolumn{4}{c|}{$\Delta[El/H]_{HD152879}$
} \\
& &
$\Delta$\Teff& 
$\Delta$\lgg& 
$\Delta$\Vt& 
$\Delta$ & &
$\Delta$\Teff& 
$\Delta$\lgg& 
$\Delta$\Vt& 
$\Delta$ \\
\hline
NA1 &  2&  0.07&  0.00& -0.02&  0.07&	  2&  0.07& -0.01& -0.05&  0.09\\
MG1 &  2&  0.05&  0.00& -0.01&  0.05&	  2& -0.03&  0.01& -0.03&  0.04\\
AL1 &  2&  0.05&  0.00& -0.01&  0.05&	  1&  0.06&  0.00& -0.03&  0.07\\
SI1 & 10& -0.02&  0.01& -0.02&  0.03&	  7& -0.10&  0.03& -0.02&  0.11\\
CA1 &  4&  0.10&  0.00& -0.04&  0.11&	  2&  0.11&  0.00& -0.04&  0.12\\
SC1 &  2&  0.14&  0.00& -0.01&  0.14&	   &      &      &      &      \\
SC2 &  4& -0.01&  0.05& -0.02&  0.05&	  3& -0.03&  0.04& -0.03&  0.06\\
TI1 & 24&  0.12& -0.01& -0.02&  0.12&	  9&  0.11&  0.00& -0.04&  0.12\\
 V1 & 26&  0.15&  0.00& -0.02&  0.15&	  1&  0.13&  0.01& -0.04&  0.14\\
CR1 &  8&  0.09&  0.00& -0.01&  0.09&	  6&  0.09&  0.01& -0.03&  0.10\\
MN1 &  1&  0.07&  0.00&  0.00&  0.07&	   &      &      &      &      \\
FE1 & 79&  0.07&  0.01& -0.02&  0.07&	 38& -0.02&  0.03& -0.03&  0.05\\
FE2 &  6& -0.08&  0.05& -0.04&  0.10&	  2& -0.22&  0.07& -0.04&  0.23\\
CO1 &  4&  0.09&  0.02&  0.00&  0.09&	  4&  0.00&  0.03& -0.02&  0.04\\
NI1 & 25&  0.05&  0.01& -0.03&  0.06&	  6& -0.04&  0.03& -0.04&  0.06\\
 Y2 &  1& -0.01&  0.05&  0.00&  0.05&	  1& -0.03&  0.03& -0.03&  0.05\\
ZR2 &  1& -0.01&  0.05&  0.00&  0.05&	   &      &      &      &      \\
BA2 &  1&  0.02&  0.03& -0.11&  0.12&	   &      &      &      &      \\
ND2 &  2&  0.01&  0.05&  0.00&  0.05&	  2&  0.01&  0.03& -0.02&  0.04\\
EU2 &  1& -0.01&  0.05& -0.01&  0.05&	  1& -0.01&  0.04& -0.02&  0.05\\
\hline
\end{tabular}
\end{table}

\begin{table*}
\captionstyle{flushleft}
\onelinecaptionsfalse 
\centering
\caption{Kinematic parameters of the investigated stars}
\label{tab:orbit}
\begin{tabular}{|c|c|c|c|c|c|c|c|c|}
\hline
N&HD    &       $U$     &       $V$     &       $W$     &
$R_{min}$ & $R_{max}$ & $Z_{max}$ & $e$ \\
   &   &        \kms     &       \kms    &        \kms   &
kpc
      & kpc       & kpc       & \\
\hline
1 &  2901&   24.4$\pm$13.0 &-144.5$\pm$9.1  & -36.9$\pm$8.9
 
& 2.17$\pm$0.16& 8.67$\pm$0.03& 0.42$\pm$0.10& 0.60$\pm$0.02\\
2 & 10550&  -57.5$\pm$3.4  & -42.1$\pm$2.7  &   3.0$\pm$2.2
 
& 6.11$\pm$0.07& 9.97$\pm$0.07& 0.36$\pm$0.00& 0.24$\pm$0.01\\
3 & 92095&   67.6$\pm$2.3  & -54.2$\pm$3.2  &  22.1$\pm$0.6
 
& 5.62$\pm$0.08& 9.42$\pm$0.04& 0.46$\pm$0.01& 0.25$\pm$0.01\\
4 & 94600&   20.5$\pm$0.6  & -18.3$\pm$0.4  & -36.9$\pm$0.4
 
& 8.17$\pm$0.01& 8.74$\pm$0.01& 0.41$\pm$0.00& 0.03$\pm$0.00\\
5 & 94669&  -41.7$\pm$0.7  & -43.3$\pm$1.6  & -38.2$\pm$0.5
 
& 6.19$\pm$0.04& 9.39$\pm$0.02& 0.45$\pm$0.00& 0.21$\pm$0.00\\
6 & 94860&   20.4$\pm$2.5  & -59.9$\pm$1.5  & -43.9$\pm$0.7
 
& 5.73$\pm$0.04& 8.62$\pm$0.01& 0.55$\pm$0.01& 0.20$\pm$0.01\\
7 & 99902&   -3.3$\pm$0.3  & -23.8$\pm$2.0  & -39.3$\pm$0.8
 
& 7.77$\pm$0.06& 8.75$\pm$0.02& 0.47$\pm$0.01& 0.06$\pm$0.00\\
8 &100696&  -40.7$\pm$0.6  & -20.4$\pm$0.3  &  35.8$\pm$0.6
 
& 7.24$\pm$0.01&10.01$\pm$0.02& 0.76$\pm$0.01& 0.16$\pm$0.00\\
9 &104985&  -76.2$\pm$1.6  &  -8.5$\pm$0.3  &  30.5$\pm$1.1
 
& 6.91$\pm$0.02&12.15$\pm$0.04& 0.74$\pm$0.02& 0.28$\pm$0.00\\
10&127243&   78.6$\pm$2.0  &-111.0$\pm$2.7  &  53.0$\pm$1.5
 
& 3.30$\pm$0.05& 9.15$\pm$0.02& 1.21$\pm$0.03& 0.47$\pm$0.01\\
11&141353&   -8.9$\pm$3.3  & -69.9$\pm$4.1  & -56.3$\pm$1.3
 
& 5.17$\pm$0.11& 8.47$\pm$0.01& 0.81$\pm$0.02& 0.24$\pm$0.01\\
12&141472&  155.0$\pm$11.7 & -71.2$\pm$5.1  &  63.7$\pm$5.1
 
& 4.10$\pm$0.10&12.74$\pm$0.37& 1.96$\pm$0.13& 0.51$\pm$0.02\\
13&152879&   10.1$\pm$0.8  & -28.5$\pm$1.6  &  58.3$\pm$2.4
 
& 7.87$\pm$0.06& 8.41$\pm$0.00& 1.22$\pm$0.04& 0.03$\pm$0.00\\
14&197752&  -78.1$\pm$3.2  &  -6.6$\pm$1.6  & -44.7$\pm$1.8
 
& 6.88$\pm$0.04&12.23$\pm$0.10& 0.67$\pm$0.02& 0.28$\pm$0.01\\
\hline
\end{tabular}
\end{table*}

\noindent
Table~\ref{tab:param} gives the parameters of the stellar atmospheres
 (effective temperature \Teff, surface gravity 
\lgg, microturbulence \Vt, and metallicity [Fe/H]), the 
stellar masses and ages on a logarithmic scale, and 
the interstellar extinctions $A_V$. We determined the 
masses and ages based on the evolutionary tracks 
from \cite{2000A&AS..141..371G} by taking into account the 
stellar metallicity. The interstellar extinctions were 
estimated from the color excesses E($B-V$),
E($V-J$), E($V-H$), E($V-K$), E($b-y$)
the normal colors were calculated from the calibrations by
\cite{1998A&A...333..231B}, which are based on Kurucz.s model 
stellar atmospheres \cite{2003IAUS..210P.A20C}. We 
determined the parameters of the stellar atmospheres 
using a technique that is also based on Kurucz.s 
model atmospheres~\cite{2003IAUS..210P.A20C} and 
analysis of the relative abundances of iron-peak elements.
 The technique is described in detail in \cite{2001ARep...45..301B} 
and allows the parameters of 
the stellar atmospheres for G-K giants to be determined
 with an accuracy of about 70-100~K for 
\Teff, 0.10-0.15 for \lgg, and 0.10-0.15~\kms\ for 
\Vt. In this paper, when analyzing the relative abundances
 of iron-peak elements when determining the 
parameters of the stellar atmospheres, we disregarded 
titanium, because it is well known that the [Ti/Fe] 
abundance can be enhanced at low metallicities and 
for thick-disk stars~\cite{2005A&A...433..185B}. Using the 
derived parameters, we computed the corresponding 
model stellar atmospheres with the ATLAS9 code 
\cite{1993KurCD..13.....K}. Based on the measured equivalent 
widths of the selected unblended spectral lines, we estimated
 the elemental abundances with the WIDTH9 
code. These are presented in Table~\ref{tab:abund} and Figure~\ref{fig:abund}, 
where the open circles and asterisks denote the abundances
 determined from the spectral lines of neutral
 and ionized atoms, respectively. The list of selected
 lines with their characteristics and equivalent 
widths is available in electronic form. The cobalt 
abundance was determined by taking into account 
the hyperfine splitting effect, which can be strong 
in the case of cool giants~\cite{2008ARep...52..630B}. 
The abundance errors given inTable~\ref{tab:abund} and marked 
by the bars in Figure~\ref{fig:abund} were determined as the dispersion
 of the individual abundances calculated from 
individual spectral lines. The possible abundance 
errors associated with the determination of stellar 
atmosphere parameters are listed in Table~\ref{tab:abnerr} for the 
stars HD~100673 (\Teff=4920~K \lgg=2.70 \Vt=1.26~\kms) and HD~152879 
(\Teff=4170K 80 \Vt=1.39~\kms) as an example. Table~\ref{tab:abnerr} gives 
the number of lines used $N$ and the changes in 
the abundance of each element when changing individual
 model parameters  ($\Delta$\Teff=+100~K, $\Delta$\lgg=+0.10,
$\Delta$\Vt=+0.10~\kms) and the total change in 
abundance.

\section*{STELLAR KINEMATICS}

\noindent
The last columns in Table~\ref{tab:list} give the membership
 probabilities of the program stars in Galactic 
subsystems whose kinematic characteristics were 
taken from \cite{2005A&A...430..165F}. We calculated the 
probabilities based on the formulas from \cite{2004A&A...418..551M}. 
Table~\ref{tab:list} includes the thick and thin disks 
and the Hercules stellar stream, because it turned 
out that some of the stars could also belong to the 
latter. In our previous paper~\cite{2011ARep...55..256P} 
devoted to the Hercules stellar stream, we showed 
that the stream is inhomogeneous in its composition 
and could contain both thin-and thick-disk stars. 
The membership probability in the halo and other 
subsystems and moving stellar groups are not given 
in Table~\ref{tab:list}, because it turned out to be less than 1\%\
for all program stars. 

Table~\ref{tab:orbit} presents the kinematic characteristics of 
the program stars. These include the Galactic velocity
 vector $(U,V,W)$ relative to the Sun and the 
Galactic orbital elements: the perigalactic distance 
$R_{min}$, the apogalactic distance $R_{max}$, the maximum 
orbital distance from the Galactic plane $Z_{max}$,the 
eccentricity $e$, and the inclination $i$. The distance to 
the Galactic center was assumed to be 8.5~kpc, while 
the necessary correction of the velocities for the solar 
motion, (+10.2, +14.9, +7.8)~\kms, was taken 
from \cite{2005A&A...430..165F}. We calculated the orbital 
elements through numerical integration of the stellar
 motion by Everhart.s 15th-order method using a 
three-component model Galactic potential~\cite{1991RMxAA..22..255A}. 
The integration accuracy was controlled
 by the conservation of the necessary integrals 
of motion. For example, in ten orbital revolutions, the 
typical relative error was $\Delta{}h/h$\textless{}$10^{-13}$ in angular 
momentum and $\Delta{}E/E$\textless{}$10^{-8}$ in total energy. The 
errors in the space velocities $(\sigma U,\sigma V,\sigma W)$ were calculated
 from the errors in the stellar proper motions, 
radial velocities, parallaxes and the errors in the solar 
velocity components relative to the local standard of 
rest. We calculated the errors in the Galactic orbital 
elements based on the model Galactic gravitational 
potential using the probable errors in the stellar space 
velocities. 

We see from Table~\ref{tab:list} that the membership probability
 in the thin disk at least for two stars (HD~94600 
and HD~99902) exceeds considerably that for the 
thick disk and these probabilities are almost equal for 
one star (HD~100696). However, these stars were 
also included in our program, because the maximum 
orbital distance from the Galactic plane $Z_{max}$ for them 
exceeds the characteristic scale height for thin-disk 
objects, 90-325 pc (\cite{1983MNRAS.202.1025G}, \cite{1996A&A...305..125R},
\cite{2001ApJ...553..184C}).

\section*{DISCUSSION}

\begin{figure*}
\centering
\captionstyle{flushleft}
\onelinecaptionsfalse 
\includegraphics[width=13cm]{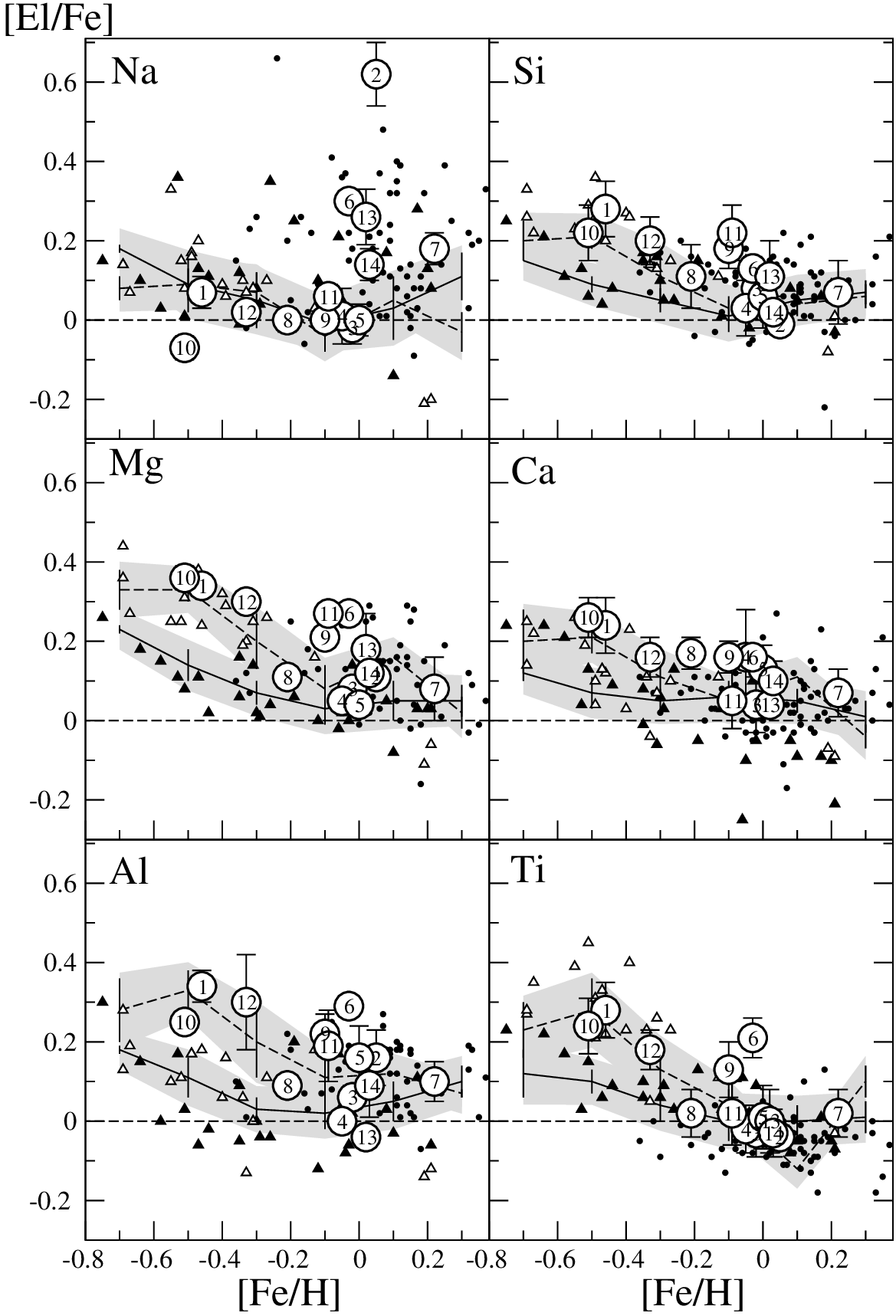}
\caption{\rm 
[Na, Mg, Al, Si, Ca, Ti/Fe] abundance trends with metallicity in the 
atmospheres of the red giants investigated here 
(the large circles with the star numbers from Table~\ref{tab:list}) in 
comparison with the thin-disk red giants that we studied previously 
(filled small circles), the thin-and thick-disk red giants from
\cite{2010A&A...513A..35A} (filled and open triangles), and the thin-
and thick-disk dwarfs from \cite{2005A&A...433..185B} (the solid and dashed lines in 
the shaded regions). 
}
\label{fig:all}
\end{figure*}

\noindent
Figure~\ref{fig:all} presents the [Na/Fe], [Mg/Fe], [Al/Fe], 
[Si/Fe], [Ca/Fe], and [Ti/Fe] abundance trends with 
metallicity in the atmospheres of the red giants investigated
 here (the large circles with ordinal numbers
 from Table~\ref{tab:list}) in comparison with other stars. 
The small filled circles indicate the 74 thin-disk red 
giants from \cite{2001ARep...45..700A, 2002ARep...46..819B,
2008ARep...52..630B, 2003ARep...47..648A, 2004ARep...48..597A,
2005ARep...49..535A, 2009ARep...53..660P, 2009ARep...53..685P,
2011ARep...55..256P} 
that we investigated by a unified technique. 
The filled and open triangles indicate the thin-disk 
(29 stars) and thick-disk (22 stars) red giants from 
\cite{2010A&A...513A..35A}, which is devoted to the 
investigation of the Galactic chemical evolution in 
the solar neighborhood. The solid and dashed lines 
indicate the thin-and thick-disk dwarfs from \cite{2005A&A...433..185B} 
averaged with a metallicity interval of 
0.2 dex, while the shaded regions denote a dispersion 
of $1\sigma$.

 The six chemical elements in Fig.~\ref{fig:all} are among 
the indicators of the Galactic chemical evolution, with 
four of them being represented by the .-elements Mg, 
Si, Ca, and Ti. We see from Fig.~\ref{fig:all} that the behavior 
of all elements, except sodium, in the atmospheres of 
the red giants is similar to that for the dwarfs. 

\begin{figure}
\captionstyle{flushleft}
\onelinecaptionsfalse 
\centering
\includegraphics[width=7cm]{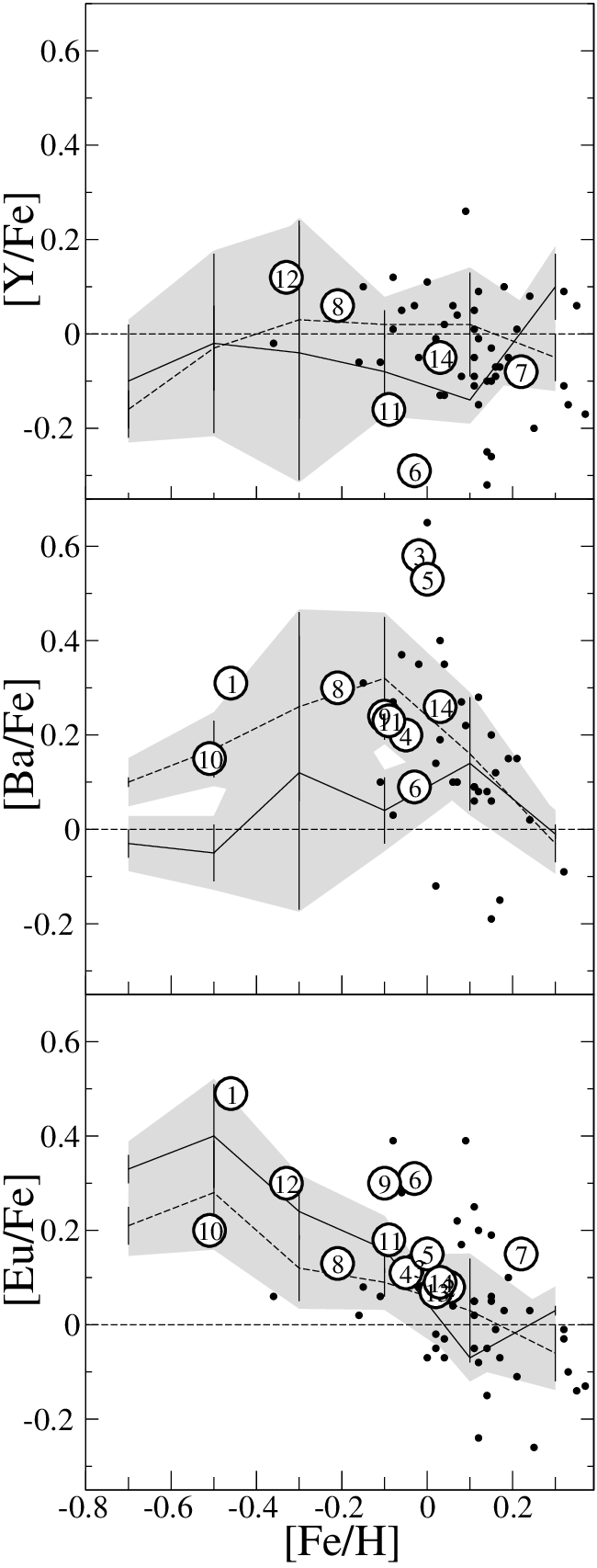}
\caption{\rm [Y, Ba, Eu/Fe] abundance trends with metallicity 
in the atmospheres of the investigated red giants in comparison
 with the data for other stars. The notation is the 
same as that in Fig.~\ref{fig:all}.}
\label{fig:all_r-s}
\end{figure}

Figure~\ref{fig:all_r-s} presents the s-element [Y/Fe] and [Ba/Fe] and 
r-element [Eu/Fe] abundance trends with metallicity. 
The designations in the figure are similar to those in Fig.~\ref{fig:all}. 
This figure does not present all stars. 
For example, the Y~II lines in some cases  are too weak for their equivalent 
widths to be properly measured, while the Ba II lines, on the contrary, often 
have an equivalent width larger than 100-150~m\AA. 
In this case, the line wings and the significantly 
growing influence of non-LTE processes play a great 
role. All of this makes it impossible to determine the 
barium abundance in this technique with an accuracy 
characteristic of other elements (about 0.1 dex).

\subsection*{Sodium} 

\noindent
The sodium abundance was determined from the 
Na~I 6154 and 6160~\AA\ lines without any correction 
for non-LTE processes. According to \cite{1989SvA....33..449K},
\cite{2000ARep...44..790M}, \cite{2011A&A...528A.103L}, 
this doublet is formed deeper than other 
sodium lines, and the non-LTE processes do not 
introduce significant corrections to the abundance 
determination. We see from Fig. 2 that a substantial 
fraction of the program stars are located above the 
dwarfs that lie compactly (the shaded region). Such 
a behavior is also typical of the previously studied 
red giants and the thin-disk giants from~\cite{2010A&A...513A..35A} 
and suggests that the [Na/Fe] abundance is determined not only by the 
chemical evolution of the Galaxy but also by the evolution of the star 
itself. 

\begin{figure}
\captionstyle{flushleft}
\onelinecaptionsfalse 
\centering
\includegraphics[width=8cm]{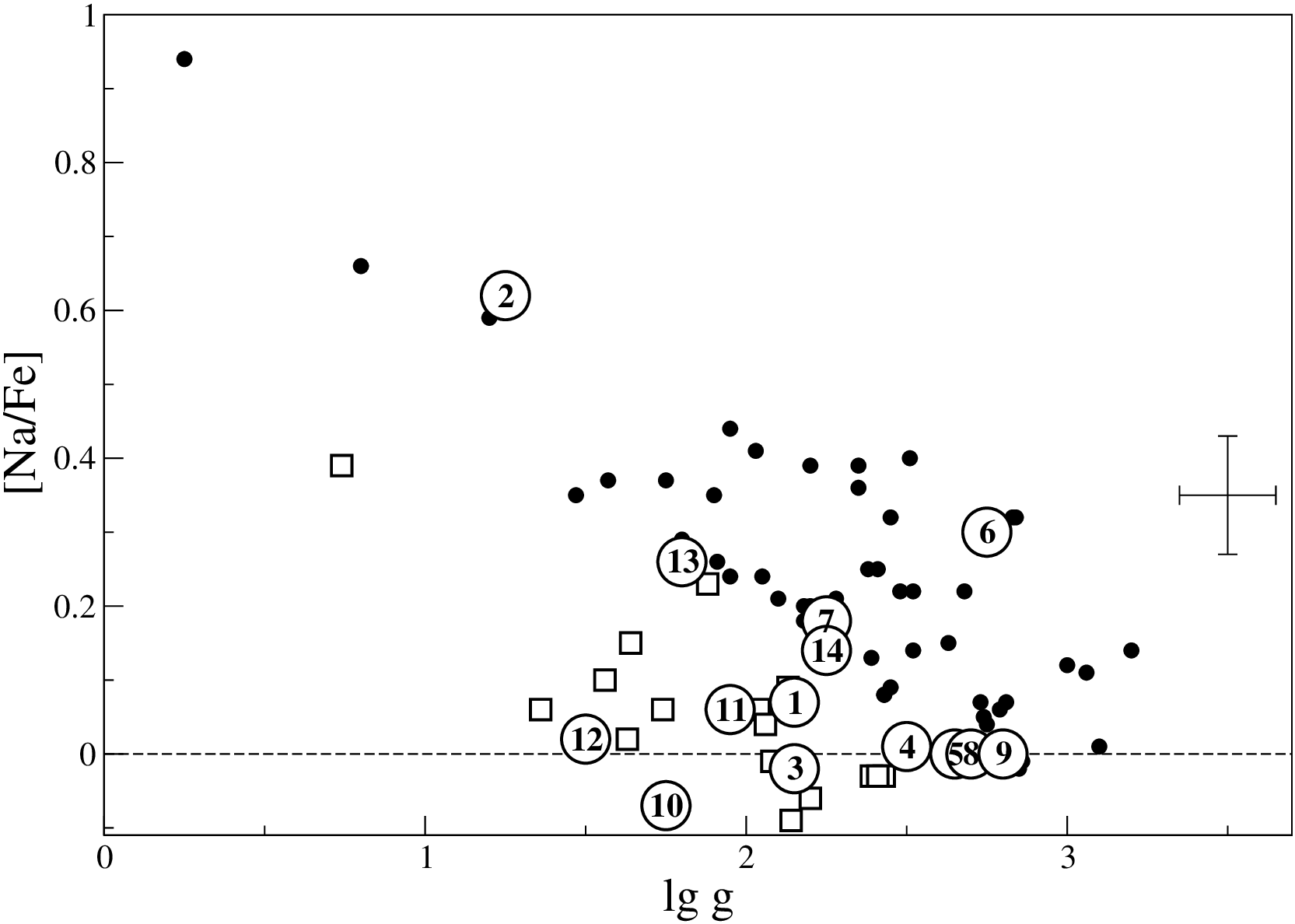}
\caption{\rm 
[Na/Fe] abundance versus surface gravity \lgg\ in the atmospheres of our red 
giants (large open circles) and the 
previously investigated thin-disk (filled circles) and thick-disk (open 
squares) red giants. The possible error bars are indicated.}
\label{fig:Na}
\end{figure}

The observed sodium overabundances in the 
atmospheres of red giants were pointed out back 
in 1963 \cite{1963ApJ...137..431C}. When 
investigating the chemical composition of the red 
giant $\epsilon~Vir$, the authors found that the abundances 
of most elements were solar, within the error limits, 
while sodium was slightly overabundant. \cite{1964ApJS....9...81H} also 
described the observed 
overabundances in the atmospheres of red giants in 
the Hyades open clyster. When investigating the 
chemical composition of red giants, the same
authors~\cite{1968ApJS...16....1H} found several stars 
with an enhanced [Na/Fe] abundance and pointed 
out a systematic difference in sodium abundance 
between red giants and dwarfs. When the results were 
discussed, it was hypothesized that sodium is synthesized
in some red giants. In 1970, \cite{1970A&A.....7..408C}
ascertained that the [Na/Fe] abundance 
increases with temperature parameter $\theta$=5040/\Teff.

The observed sodium overabundances in the 
atmospheres of supergiants were detected in 1981 
\cite{1981BCrAO..63...68B, 1981BCrAO..64....1B}. 
It was then also shown 
that the overabundance depends on surface gravity 
\lgg\ or luminosity. Subsequently, these authors 
hypothesized that sodium could be produced in the 
$^{22}$Ne(p,$\gamma$)$^{23}$Na reaction that enters the neon-sodium 
cycle of hydrogen burning in the cores of main-sequence 
stars and could then be brought from deep 
layers into the stellar atmosphere through developing
 convection as the star evolves from the main 
sequence to the red giant branch~\cite{1983BCrAO..66..119B}. 
The calculations performed by \cite{1987SvAL...13..214D, 1988SvAL...14..435D} 
confirmed that this hypothesis is possible. Subsequent
calculations using new nuclear reaction 
characteristics also showed a change of the sodium 
abundance in red giants~\cite{1996ApJ...464L..79C, 1998ApJ...492..575C}. 
Although the neon-sodium cycle requires high temperatures
and is efficient only in massive stars, low-mass (1-3 $M_\odot$) stars 
can also produce sodium. 

The stable isotope $^{23}Na$ is currently believed to 
be produced mainly in the carbon burning reaction 
$^{12}C(^{12}C,p)^{23}Na$. A small amount can also be produced
in the hydrogen, $^{22}Ne(p,\gamma)^{23}Na$, and helium, 
$^{14}N(\alpha,\gamma)^{19}F(\alpha,\gamma)^{23}Na$, burning reactions 
\cite{1995ApJS..101..181W}. The first and last reactions do not 
proceed in red giants at the shell hydrogen burning 
stage, but the second reaction is possible. Thus, the 
helium and carbon burning reactions in stars of earlier
 generations for red giants determine some basic 
sodium abundance, as in the case of dwarfs, and the 
amount of sodium produced in the stellar interior over 
its lifetime on the main sequence in the hydrogen 
burning reaction is added to this value. According 
to the calculations published in the review~\cite{RevModPhys..69.995W}, 
the $^{23}Na$ abundance at a temperature T=$0.05\cdot10^9$~K 
and a pressure of $100~g cm^{-3}$ the stellar core increases by a factor 
of 7 in 100 yr. 
However, there are large uncertainties in the data on 
the $^{22}Ne(p,\gamma)^{23}Na$ reaction cross sections and rates 
under conditions of low temperatures in the stellar 
cores. 

Subsequently, it was shown that the observed 
[Na/Fe] overabundances in the atmospheres of normal
 red giants also depend on surface gravity~\cite{2001ARep...45..301B}, 
with this dependence being an 
extension of the dependence for supergiants noted 
above. Interestingly, [Na/Fe] overabundances were 
also detected in the atmospheres of various classes 
of red giants: moderate and classical barium stars 
\cite{2002ARep...46..819B, 2004ARep...48..597A} supermetal-
rich stars~\cite{2009ARep...53..660P, 2009ARep...53..685P}, 
whose positions satisfy the same dependence as that 
for the remaining normal red giants. The subsequent 
sodium abundance analyses for the atmospheres of 
red giants performed at the Institute of Astronomy, 
the Russian Academy of Sciences, showed that some 
stars have a reduced [Na/Fe] abundance relative to 
the observed dependence on surface gravity~\cite{2005ARep...49..535A,
2009ARep...53..660P, 2009ARep...53..685P}. The 
kinematic characteristics of such stars correspond to 
thick-disk objects. 

In Fig.~\ref{fig:all}, HD~10550 lies above the remaining 
stars and has a sodium abundance that is higher 
than that for dwarfs at the same metallicity approximately
 by 0.5~dex. Among all of our program stars, 
HD~10550 has the lowest surface gravity \lgg=1.25. 
HD~152879 also lies statistically above the 
dwarfs (\lgg=1.80). The remaining stars lie, within 
the error limits, in the region where the dwarfs are 
located, except HD~127243; it has a [Na/Fe] abundance
that is lower than that for dwarfs approximately 
by 0.15~dex. 

In Fig.~\ref{fig:Na}, the [Na/Fe] abundance is plotted against 
the surface gravity \lgg\ in the atmospheres of the 
program stars in comparison with the previously 
studied thin-and thick-disk red giants. We see that 
the positions of the thin-disk stars form a distinct 
dependence, while the thick-disk stars with a reduced 
[Na/Fe] abundance lie below it. In this figure, most of 
the program stars are also located below the dependence
 for the thin-disk giants in the region where the 
thick-disk stars dominate. For some of the stars with 
a relatively low metallicity (HD~127243 [Fe/H]=-0.44
and HD~141472 [Fe/H]=-0.27), the [Na/Fe] 
underabundance reaches 0.5~dex. The stars from 
Table~\ref{tab:list} with a high probability of their membership 
in the thin disk but with a large orbital parameter 
$Z_{max}$ (HD~94600, HD~99902, and HD~100696) are 
located at the lower boundary of the region for the 
thin-disk giants and at its upper boundary for the 
thick-disk giants. In this case, we cannot determine 
with confidence which dynamical subsystem of the 
Galaxy they belong to from Fig.~\ref{fig:Na}. 

The causes of the reduced sodium abundance in 
the atmospheres of thick-disk red giants relative to 
the [Na/Fe]-\lgg\ relation can be: (1) the initial 
sodium underabundance; (2) the initial neon under-
abundance that led to the production of a smaller 
amount of sodium; and (3) the shallower convective 
shell. The first assumption is not backed by the observational
 data; the [Na/Fe] abundances in the thick-
and thin-disk dwarfs differ only slightly. The second 
assumption can possibly play a significant role, but 
it is difficult to test this hypothesis at present due 
to the absence of accurate data on the neon abundances
 in thick-disk stars. The last assumption is 
unlikely, because the physical parameters of the stars 
(\Teff, \lgg, [Fe/H], mass) with normal and reduced 
[Na/Fe] abundances are identical and it would be 
unreasonable to expect the convective processes in 
them to differ noticeably. 

Among the program thick-disk stars, there are 
two with a significant [Na/Fe] abundance relative to 
other stars with similar parameters. The first star 
HD~94860 is located in a region characteristic of 
the metal-rick thin-disk stars. For example, the 
Hyades giants (with [Fe/H]$\approx$0.14~dex) are also located
there. The equivalent widths of the sodium 
6154 and 6160~\AA\ lines are 85 and 105~m\AA, respectively;
such very close values were also measured for 
HD~94669 with similar \lgg\ and [Fe/H], but this star 
has a much lower temperature (by 250~K) and, hence, 
a low abundance [Na/Fe]=0.00~dex. Since the errors
 in the [Na/Fe] abundance (see Table~\ref{tab:abnerr}) cannot 
change it greatly, the estimated sodium abundance in 
HD~94860 is realistic. This star, which kinematically 
belongs to the thick disk, may have been formed in 
agas.dust cloud similar in chemical composition to 
the thin-disk objects. The second star HD~10500 
is located in Fig.~\ref{fig:Na} near the thin-disk supergiants. 
However, it has the highest mass among the program 
stars (7$M_\odot$) and, consequently, the internal physical 
conditions under which sodium is synthesized more 
efficiently. The star may also have a deeper convective 
envelope that provides the rise of sodium from the 
stellar interior. 

\subsection*{Magnesium}

In Fig.~\ref{fig:all}, all of the program stars are located, 
within the magnesium abundance error limits, in the 
region of dwarfs. Three red giants with the lowest iron 
abundance (HD~127243, HD~2901, and HD~141472) 
clearly correspond to the distribution of thick-disk 
stars. In this region with [Fe/H]$\lesssim$-0.3, the separation
 of the thin-and thick-disk dwarfs is largest 
(about 0.2~dex) and exceeds the mean abundance 
errors (about 0.1~dex). The data on the giants from 
\cite{2010A&A...513A..35A} also closely correspond to 
the trends for the dwarfs: the thin-disk giants lie 
along the solid line, while the thick-disk giants lie 
along the dashed line. 

Magnesium is produced in the carbon, 
$^{16}O(^{12}C,\alpha)^{24}Mg$, helium, $^{20}Ne(\alpha,\gamma)^{24}Mg$, 
$^{21}Ne(\alpha,n)^{24}Mg$, and hydrogen, $^{23}Na(p,\gamma)^{24}Mg$, 
$^{27}Al(p,\alpha)^{24}Mg$, burning reactions. The case of carbon
 and helium burning plays no role in the chemical 
evolution during the lifetime of red giants, while the 
last hydrogen burning reaction in the magnesium. 
aluminum cycle is mots likely inefficient, because 
this requires conditions with much higher temperatures
 than those in low-mass stars at the red giant 
stage. Only the reaction with sodium involvement 
$^{23}Na(p,\gamma)^{24}Mg$ can slightly raise the magnesium 
abundance \cite{RevModPhys..69.995W}. This may 
underlie the slight [Mg/Fe] overabundance for some 
normal red giants. However, the entire effect is within 
the $2\sigma$ magnesium abundance error limits. 

\subsection*{Aluminum}

In Fig.~\ref{fig:all}, all of the program stars are located, 
within the aluminum abundance error limits, in the 
region of dwarfs. Just as in the case of magnesium,
 three red giants with the lowest iron abundance 
(HD~127243, HD~2901, and HD~141472) clearly 
correspond to the distribution of thick-disk dwarfs. 
The data from~\cite{2010A&A...513A..35A} show a larger 
scatter of abundances relative to the dwarfs than that 
for magnesium. Some of the previously studied red 
giants are located approximately 0.2~dex higher than 
the dwarfs. Although this difference does not exceed 
the $2\sigma$ level, the published data suggest an enhanced 
aluminum abundance in globular cluster giants~\cite{1996ApJ...464L..79C}. 

In stars, aluminum can be produced 
in two reactions of the magnesium.aluminum cycle 
of hydrogen burning: $^{26}Mg(p,\gamma)^{27}Al$ and 
$^{26}Al(p,\gamma)^{27}Si(\beta^+\nu_e)^{27}Al$. The physical parameters 
needed for the $^{27}Al$ production reaction were estimated
 by~\cite{1999ApJ...519..733F}, who described the 
possibility of the existence of observable aluminum 
overabundances in the atmospheres of red giants. 

\subsection*{Silicon}
\noindent

In Fig.~\ref{fig:all}, all of the program stars are located 
within the silicon abundance error limits, in the region
 of dwarfs. Just as in the case of magnesium 
and aluminum, three red giants with the lowest iron 
abundance (HD~127243, HD~2901, and HD~141472) 
correspond to the distribution of thick-disk dwarfs. 

The isotope $^{28}Si$  is produced mainly in the oxygen 
burning reaction $^{16}O(^{16}O,\alpha)^{28}Si$, which is inaccessible
 for low-mass red giants. The $^{27}Al(p,\gamma)^{28}Si$ 
reaction entering the magnesium.aluminum cycle of 
hydrogen burning can be a possible silicon production
 source, but its contribution to nucleosynthesis 
in low-mass stars is negligible and is considered as 
a leakage reaction~\cite{1996ApJ...464L..79C}. 

\subsection*{Calcium}
\noindent

In Fig.~\ref{fig:all}, almost all of the program stars are 
located, within the calcium abundance error limits, in 
the region of dwarfs. Just as in the case of magnesium 
and aluminum, three red giants with the lowest iron 
abundance (HD~127243, HD~2901, and HD~141472) 
correspond to the distribution of thick-disk dwarfs. 

Calcium is an $\alpha$-element and it originates in the 
oxygen burning reactions. The calcium abundance 
does not change over the lifetime of a red giant. 

\subsection*{Titanium}
\noindent

Titanium, along with magnesium, silicon, and 
calcium, is an $\alpha$-element and, at the same time, 
also belongs to the iron-peak elements produced 
during supernova 1a explosions. Just as other .elements,
 titan is characterized by an increase in its 
abundance with decreasing metallicity. Just as in the 
case of magnesium, aluminum, silicon, and calcium, 
three red giants with the lowest iron abundance 
(HD~127243, HD~2901, and HD~141472) correspond 
to the distribution of thick-disk dwarfs. A slightly 
overestimated titanium abundance was determined 
for HD 94860, but it is within the $1\sigma$ error limits 
relative to the dwarfs.

\subsection*{Yttrium}
\noindent

As one of the s-elements, yttrium is produced in 
the slow neutron capture process. The presence of 
a flux of free neutrons is necessary for this process. 
This condition is met in stars at the asymptotic giant 
branch stage.a later evolutionary stage relative to 
the normal red giants. Therefore, the same picture as 
that for the dwarf stars must be observed (Fig.~\ref{fig:all_r-s}). 

However, this figure does not present the moderate 
and classical barium stars investigated previously 
\cite{2002ARep...46..819B, 2004ARep...48..597A}. In 
classical barium stars, binarity plays a major role. 
one of the binary components evolves more rapidly 
and ejects its envelope rich in s-elements part of 
which is captured by the secondary component. 
They are characterized by a different, steeper trend 
with decreasing metallicity~\cite{2004ARep...48..597A}. 
In contrast, for moderate barium stars, some of 
which can be represented by single red giants, possible
 photoneutron reactions like $^{13}C(\gamma,n)^{12}C$ and 
$^{14}N(\gamma,n)^{13}N$ can play a role. According to
\cite{1974ApJ...187..303H}, when the CNO cycle of 
hydrogen burning proceeds in the interior of a main-
sequence star, a large number of energetic gamma-
ray photons capable of knocking out a neutron from 
the nuclei of some elements emerge. This source of 
neutrons is weak, but, as was pointed out by \cite{1974ApJ...187..303H}, 
it is the second in intensity 
among all of the known neutron sources and could 
possibly explain the small anomalies of s-process 
elements observed in stars at the red giant stage. 

\subsection*{Barium}
\noindent

In Fig.~\ref{fig:all_r-s}, all of the program stars, except two, 
lie in the region of dwarfs. These stars (HD~92095 
and HD~94669) have the most saturated barium lines 
among the stars presented on the plot and probably 
the largest errors in the barium abundance. The 
most metal-poor stars HD~127243 and HD~2901 are 
located in a region characteristic of the thick-disk 
dwarfs. 

\subsection*{Europium}
\noindent

Europium is an r-element whose origin is associated
 with the rapid neutron capture process that 
occurs during supernova explosions, and its abundance
 does not change during the evolution of a red 
giant. Within the error limits, the program stars are 
located in the same region as the dwarfs and exhibit 
an increase in [Eu/Fe] abundance with decreasing 
metallicity. 

\section*{CONCLUSIONS}

We investigated 14 red giants with high space 
velocities. For all stars, we determined their Galactic 
orbital elements and other kinematic characteristics 
from which their membership in a particular Galactic 
subsystem can be judged. We determined the physical
 parameters, stellar atmosphere parameters, and 
elemental abundances. For almost all chemical elements,
 their abundance trends with metallicity were 
shown to correspond to those for dwarfs. Where a 
significant difference exists between the trends corresponding
 to the thin-and thick-disk dwarfs, the 
positions of the thick-disk giants follow the trend 
for the thick-disk dwarfs. An abundance analysis of 
the atmospheres of red giants provides an additional 
possibility to determine whether a star belongs to the 
Galactic thick or thin disk in some cases. A different 
picture is observed for sodium: in addition to the 
dependence on metallicity, which is determined by the 
chemical evolution of the Galaxy, there is a significant 
dependence on surface gravity \lgg, which reflects 
the degree of development of the convective envelope 
in a red giant. In this case, there is a significant 
difference between the [Na/Fe] abundances in the 
atmospheres of thick-and thin-disk red giants. 

\textit{Acknowledgments}
\noindent
This work was supported by the Russian Foundation
 for Basic Research (projects no. 09-02-00528-a, 
12-02-00610-a), the ''Origin, Structure, and Evolution
 of Objects in the Universe''. Program of the Presidium
 of the Russian Academy of Sciences, and the 
Federal Agency for Science and Innovations (State 
contract no. 02.740.11.0247). 

\vskip 5mm
\bibliographystyle{aipauth4-1} 
\bibliography{paper}

\end{document}